\documentclass[aps,prl,reprint,showpacs,floatfix]{revtex4-1}
\usepackage{amsmath}
\usepackage{amssymb}
\usepackage{graphicx}
\usepackage[normalem]{ulem}

\begin{document}

\title{Measurement induced chaos with entangled states}

\author{T. Kiss$^{1}$, S. Vym\v etal$^{2}$, L. D. T\'oth$^{1,3}$, A. G\'abris$^{1,2}$, I. Jex$^{2}$, G. Alber$^{4}$}
\affiliation{
$^{1}$Research Institute for Solid State Physics and Optics, HAS,  P.O.B. 49, H-1121 Budapest, Hungary \\
$^{2}$%Department of Physics, 
Faculty of Nuclear Sciences and Physical Engineering, 
Czech Technical University in Prague, B{\v r}ehov\'a 7, 115 19 Praha 1 - Star\'e M\v{e}sto, Czech Republic \\
$^{3}$DAMTP, University of Cambridge, UK \\
$^{4}$IAP, TU Darmstadt,
D-64289 Darmstadt, Germany}

\date{\today}

\begin{abstract} 
  The dynamics of an ensemble of identically prepared two-qubit
  systems is investigated which is subjected to the iteratively
  applied measurements and conditional selection of a typical
  entanglement purification protocol.  It is shown that the resulting
  measurement-induced non-linear dynamics of the two-qubit state
  exhibits strong sensitivity to initial conditions and also true
  chaos. For a special class of initially prepared two-qubit states
  two types of islands characterize the asymptotic limit. They
  correspond to a separable and a maximally entangled two-qubit state,
  respectively, and their boundaries form fractal-like structures. In
  the presence of incoherent noise an additional stable asymptotic
  cycle appears.
\end{abstract}

\maketitle

{\em Introduction---}Entanglement is at the heart of quantum physics
since its discovery \cite{Verschrankung}. However, it was only
recently that the focus has been put on entanglement as a resource for
quantum communication and quantum information processing
\cite{Horodecki}. Various protocols have been developed to detect
\cite{Guhne}, generate and distill entanglement
\cite{PurificationReview}. From an ensemble of identical quantum
states, one can produce an ensemble yielding higher degree of
entanglement by unitary transformations, measurements and selection
conditioned on measurement outcomes. These non-linear processes are
referred to as entanglement distillation or purification. Entanglement
distillation protocols play a crucial role in increasing the quality
of communication channels and have also been used to define the degree
of entanglement in an operational sense.

The phenomenon that sensitivity to initial conditions leads to chaotic
dynamics in classical physics is well-known. Similar phenomena in
closed quantum systems are, however, excluded by the quantum unitary
evolution \cite{Chaos}. In the case of open quantum systems the
restriction to unitarity is lifted.  The evolution of an open quantum
system is sometimes pictured as additionally having an environment that
performs generalized measurements on it. In general, any type of measurement
makes the evolution non-unitary. 
Entanglement purification, i.e. selection of certain systems from an ensemble of identical
systems based on the results of partial measurements, can be regarded
as a generalized feedback mechanism. The non-linear dynamics resulting
from such generalized feedback has been shown to lead in certain cases
to a chaos \cite{ChaoticQubit}. This type of chaos is essentially
different from that arising from the stochastic dynamics of a
continuously measured open quantum system \cite{quantum-classical},
which can become chaotic in the semi-classical regime while still
showing signatures of quantum behaviour \cite{HJS}.

The generalized feedback resulting from measurement based selection
plays a crucial role in the case of various entanglement purification
protocols \cite{PurificationReview}. 
%For simplicity it is usually
%assumed that the quantum systems in the ensemble are prepared
%initially in identical quantum states, and after each iteration of the
%purification process those systems which are retained are again in
%identical quantum states. Thus, although the physical realization of
%the protocol involves the processing of an ensemble of identical
%quantum systems, the mathematical description can be given using only
%a quantum state of a single instance. 
Applying the purification protocol \cite{Gisin98}  on an
ensemble of single qubits prepared in identical pure states, 
after each iteration step the remaining, selected ensemble of qubits  will again be in identical, pure quantum states. The
dynamical evolution can be characterized by a rational non-linear map over
the extended complex plane representing the pure states. 
This map has been proven to lead to truly
chaotic behaviour \cite{ChaoticQubit}. A single qubit is the simplest
quantum system, and also lacks all genuine non-local quantum
properties. Thus the above mentioned study left a fundamental question
open, which in turn raises naturally in the context of entanglement
purification: Can we find sensitivity to initial conditions for
genuine mulipartite quantum properties, in particular for
entanglement, when only local operations and classical communication
(LOCC) are applied?

In this Letter we focus on a particular entanglement purification
protocol \cite{Gisin98}, and demonstrate the existence of true chaos
which manifests also in the evolution of entanglement. An important
feature of the protocol is that it maps pure states onto pure states,
moreover it may also increase purity of initially mixed states. Our
analysis is accomplished by showing that the convergence to a fully
entangled or a separable asymptotic attractor can be sensitive to the
initial state.

\begin{figure}
\begin{center}
\includegraphics[width=0.42 \textwidth]{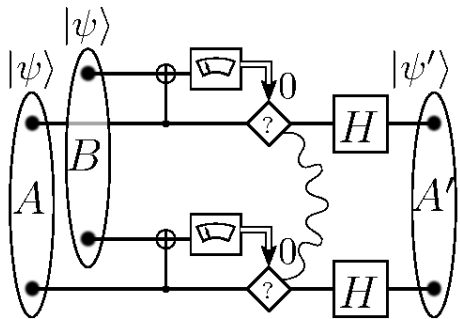}
\caption{Schematic representation of one iteration step of the
  entanglement purification process. The diamond shaped elements
  denote acceptance of the pair of qubit only when both measurements
  yield 0. The depicted procedure is used to prepare from an ensemble
  $\mathcal{E}$ made up of pairs of qubits all in a state
  $|\psi\rangle$, an ensemble $\mathcal{E'}$ made up of pairs of
  qubits in the state $|\psi'\rangle$. During the procedure half of
  the pairs in $\mathcal{E}$ are completely used up, while a portion
  of the other half is retained\ depending on the measurement
  outcome. Here we depicted the special case $U = H \otimes H$, where
  $H$ is the Hadamard gate. }
\label{fig1}
\end{center}
\end{figure}

{\em Measurement based non-linear dynamics}--- The pure quantum state
state of a system consisting of a pair of qubits can be expressed in
the computational basis as
$|\Psi\rangle=c_1|00\rangle+c_2|01\rangle+c_3|10\rangle + c_4
|11\rangle$. We consider an ensemble of qubit pairs prepared in the
state $|\Psi\rangle$ as the input to the entanglement purification
protocol \cite{Gisin98} schematically depicted on
Fig.~\ref{fig1}. This protocol realizes a non-linear transformation of
the quantum state $|\Psi\rangle$ according to $c_{i}
\overset{S}{\longrightarrow} N c_{i}^2$, where $N$ is a necessary
normalization factor. The non-linearity is due to the generalized
feedback realized by the measurement-based conditional selection. By
following each non-linear transformation by a unitary $U$, further
non-trivial dynamics can be generated. A complete iteration can be
expressed as the transformation
%%%
\begin{equation}
|\Psi^{\prime}\rangle =  U S |\Psi \rangle\,.
\end{equation}
%%%
Note, that only local unitary transformations comply with the concept
of entanglement purification. For the following analysis we fix the
unitary transformation to be $U=H\otimes H$, where $H$ is a one qubit
Hadamard gate ($H_{ij}=(-1)^{i\cdot j}/\sqrt{2}$).

For Hilbert spaces of dimensionalities of more than two, such as for
that of a two qubit system, the representation of a pure state
requires a vector of several complex numbers, and the non-linear
dynamics is described by a non-linear map on this higher dimensional
complex space. The mathematics of non-linear dynamical maps in several
complex variables is substantially more involved and much less
understood than the same for a single complex variable. Even the
existence of chaotic regions is a nontrivial question \cite{Fornaess}.

We are interested in the evolution of genuine multi-partite properties
of the system under the iterative non-linear dynamics. Since this
protocol describes an entanglement purification protocol, of all such
properties, entanglement is of the most concern. Thus we consider
states parametrized by the complex number including infinity $\zeta$
($\zeta \in \hat{\mathbb{C}}$) of the form
%%%%%%%
\begin{equation}
\label{eq:initial}
|\Psi(\zeta)\rangle={\cal N}(\zeta)
\left( 
|00\rangle+
\zeta|11\rangle
\right) \, ,
\end{equation}
%%%%%%%
where the normalisation factor reads ${\cal
  N}(\zeta)=(1+|\zeta|^2)^{-1/2}$. The degree of entanglement of this
state is completely determined by $\zeta$ via the binary entropy
function $H({\cal N}(\zeta)^2)$ \cite{SchmidtBasisNote}.

By studying the evolution of these states we can learn about
the general multi-partite properties of this protocol. The analytic
treatment of this evolution is greatly simplified by the fact that the
states of the form of Eq.~(\ref{eq:initial}) are invariant under two
successive iterations, in particular yielding
%%%%%%%
\begin{equation}
|\Psi(\zeta)^{(2)}\rangle={\cal N}(g(\zeta))
\left( 
|00\rangle+g(\zeta)|11\rangle
\right) \, ,
\end{equation}
%%%%%%%
where 
%%%%%%%
\begin{equation}
g(\zeta)=\frac{2 \zeta^2}{1+\zeta^4}
 \, .
\end{equation}
%%%%%%%
Thus the description of the dynamics for this class of initial states
simplifies to a non-linear map of a single complex variable. The
function $g(\zeta): \hat{\mathbb{C}}\mapsto\hat{\mathbb{C}}$ is a
fourth order rational function generating a one variable complex
dynamical map on the Riemann sphere. The map is chaotic in the sense
that the corresponding Julia set is non-vacuous \cite{Milnor}. Fourth
order maps in general can lead to rather involved behaviour. We base
our analysis on the observation that the fourth order rational map can
be written as a composition of a second order rational function with
itself $g(\zeta)=f(f(\zeta))=f^{\circ 2}(\zeta)$, where
%%%%%%%
\begin{equation}
\label{eq:fmap}
f(\zeta)=\frac{1- \zeta^2}{1+\zeta^2}
 \, .
\end{equation}
%%%%%%%
Moreover, the same function also describes the quantum state after an
odd number of iterations, according to
%%%%%%%
\begin{equation}
\label{eq:odd}
|\Psi(\zeta)^{(2n+1)}\rangle={\cal N}
\left( 
  |00\rangle+|11\rangle
+f^{\circ 2n+1}(\zeta) (|01\rangle+|10\rangle)
\right).
\end{equation}
%%%%%%%
Thus the iterative dynamics of initial states of the form
Eq.~(\ref{eq:initial}) is equivalent to the iterative dynamics
generated by the second order rational map defined in
Eq.~(\ref{eq:fmap}). In the following we will examine this mapping in
some detail.

{\em Stable cycles of the dynamics---}The long term behaviour of a rational map can be analyzed \cite{Milnor} by following the orbits of its critical points (points where the derivative of the map vanishes $f^\prime(\zeta_{c})=0$). 
The second order map in Eq.~(\ref{eq:fmap}) has two critical ponts: $\zeta_{c1}=0$ and $\zeta_{c2}=\infty$.
The first one, $\zeta_{c1}$ is part of the superattracting cycle $\{0,1\}$, while the second one lands on the same cycle after two iterations. Therefore, the map has one stable (superattractive) cycle. 
Translated back to the language of states, we have to take into account that the function $f$ describes a state in the same form only in every second step. Thus, depending on the parity of steps when reaching the first element of the cycle, we have two distinct cases.  

In the first case, when $f^{\circ 2n}(\zeta)=1$, the corresponding state reads 
%%%%%%%
\begin{equation}
\label{eq:bell}
|\Psi^{(2n)}\rangle=
\frac{1}{\sqrt{2}}
\left( 
|00\rangle+|11\rangle
\right)=|\Phi^{+}\rangle \, .
\end{equation}
%%%%%%%
After an even number of steps, one reaches a fully entangled state, the Bell state $|\Phi^{+}\rangle$. Since this state is invariant under both the non-linear transformation $S$ and the unitary operation $H \otimes H$, we have $|\Psi^{(2n)}\rangle=|\Psi^{(2n+1)}\rangle=|\Phi^{+}\rangle $. Therefore, the Bell state $|\Phi^{+}\rangle$ is an asymptotically stable fixed point of the dynamics.

In the second case, when $f^{\circ 2n}(\zeta)=0$, the corresponding state reads
%%%%%%%
\begin{equation}
\label{eq:separable1}
|\Psi^{(2n)}\rangle= |00\rangle \, .
\end{equation}
%%%%%%%
The above product state is not invariant under the iterative dynamics,
however, any subsequent step will leave the state completely
separable. In particular, after an odd number of steps we find
%%%%%%%
\begin{equation}
\label{eq:separable2}
|\Psi^{(2n+1)}\rangle= \frac{1}{2} \left( |0\rangle + |1\rangle \right) \otimes \left(|0\rangle+|1\rangle\right) \, .
\end{equation}
%%%%%%%
Therefore, the second stable cycle is of length two, both of its members are separable pure states.

Due to a theorem on rational maps \cite{Milnor}, the degree of the rational function determines the maximum number of stable cycles. For a rational map of degree two, at most two stable cycles can exist. In our case, the single stable cycle of the rational map $f$ can lead to two different stable cycles of the dynamics, depending on the parity of the number of steps when approaching the first element of the cycle. Thus, we have found all possible stable cycles of the dynamics restricted to initial pure states of the form (\ref{eq:initial}).

{\em Sensitivity to initial states---}The two possible stable cycles of the dynamics are very different. One of them is a single, completely entangled pure state, a Bell state, while the other is an oscillation between two  separable pure states of the two qubits.
We ask now the question, what are the initial states converging to each of the stable cycles. 

Let us first discuss the case of real values for the parameter
$\zeta$. The function $f$ maps real numbers to real numbers, thus it
can be restricted to $\mathbb{R}$. The members of the stable cycle
$\{0,1\}$ are also real numbers.  We can now determine the basin of
attraction for the two cases of convergence, i.e.\ convergence to $0$
after an even or an odd number of steps, which we shall call even-zero
or odd-zero convergence, respectively. The immediate neighbourhood of
the fixed point $0$, belonging to even-zero convergence, is determined
by the equation $|f^{\circ 2}(\zeta)|<|\zeta|$, with the condition
$|\zeta|<1$. The corresponding equation can be explicitly solved
yielding $|\zeta|<\zeta_{A}$ where
\begin{equation}
\zeta_{A}=(a-1-2/a)/3 \quad
 \mathrm{with} \quad a=(17+3 \sqrt{33})^{1/3} \, .
\end{equation}
The preimages of the interval $(-\zeta_{A},\zeta_{A})$ belong to an odd-zero convergence region. By solving the corresponding equation we find two distinct intervals of odd-zero convergence
$\zeta_{A}<|\zeta|<\zeta_{B}$ where
\begin{equation}
\zeta_{B}= \sqrt{\frac{-2+2 a+ a^{2}}{2+4 a-a^{2}}} \, .
\end{equation}
It is easy to see that the region $|\zeta|>\zeta_{B}$ is mapped after one iteration to the region $\zeta_{A}<|\zeta|<\zeta_{B}$, thus it belongs to even-zero convergence. To summarize the behaviour of the map restricted to the reals, we have found that regions of odd-zero and even-zero convergence follow each other, these open sets belong to the Fatou set. The border point $\zeta_{A}$ is a repulsive fixed point of the map, while the other border points $\{ -\zeta_{A},-\zeta_{B},\zeta_{B}\}$ are preimages of $\zeta_{A}$. These four points belong to the Julia set.

\begin{figure}
\begin{center}
\includegraphics[width=0.42 \textwidth]{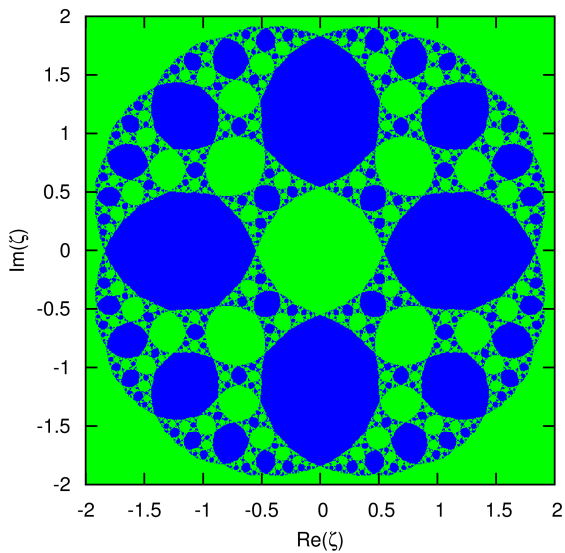}
\caption{Convergence of initial states of Eq.~(\ref{eq:initial}) to
  the two limiting cycles as a function of the complex parameter
  $\zeta$. The blue color denotes convergence to the maximally
  entangled state of Eq.~(\ref{eq:bell}), and the green color the
  convergence to the cycle of the two separable states in
  Eqs. (\ref{eq:separable1}) and (\ref{eq:separable2}).}
\label{fig2}
\end{center}
\end{figure}

A similar analysis can be repeated for the map $f$ with domain $\hat{\mathbb{C}}$. The preimages of $0$ with their open small neighbourhoods provide regions of odd-zero or even-zero convergence, forming the Fatou set. The  Fatou set is an open set, while the complementary closed set is the Julia set. A small neighbourhood of the origin will belong to  even-zero convergence, a sufficient condition for this is that $|f(f(\zeta))|<|\zeta|$. From this condition we can determine the maximum radius $\zeta_{C}$ of a circle around the origin belonging to even-zero convergence. The radius $\zeta_{C}$ can be calculated explicitly from an algebraic equation, its numerical value is $\zeta_{C}\approx 0.475$. The first order pre-images of $0$ are $1$ and $-1$. Thus, $1$ and $-1$ together with a region around them belong to odd-zero convergence. The preimage of the circle of convergence around zero determines an immediate region of convergence around $1$ and $-1$. In a similar manner one can continue this process and determine the next order preimages and regions of convergence around them. The numerically calculated convergence regions are shown in Fig. \ref{fig2}.
We can clearly recognize a chessboard-like structure of green and blue islands, belonging to convergence to fully entangled or completely separable states, respectively. The islands follow a self similar structure with decreasing size.  A Julia set is formed by the border points between the two colors, with no stable cycles in it.
The asymptotic entanglement of the two-qubit system behaves chaotically, it is sensitive to the initial state on arbitrary small scales. Since this is an asymptotic map, it involves an infinite-in-time limit process.

{\em Mixed initial states---}Up to now, we have considered pure initial states. Adding noise to the initial state, described by a density operator, will
alter the dynamics. We will test the sensitivity to a certain type of noise by adding the unit matrix to the density matrix representing the initial pure state
%%%%%%%
\begin{equation}
\label{eq:Werner}
\rho(\zeta,\lambda)= \lambda |\Psi(\zeta)\rangle \langle\Psi(\zeta)|+ \frac{1-\lambda}{4} \openone \, ,
\end{equation}
%%%%%%%
where $\lambda \in \left[ 0,1\right]$, and $\openone$ stands for the
unit operator acting on the Hilbert space of the two qubits. Since the
dynamics is no longer restricted to pure states, we can expect that
further stable fixed cycles will appear, containing mixed states. The stability of the fixed cycles can be proven in any convenient representation. We chose the Fano representation where the real expansion coefficients with respect to the 16 generalized Pauli matrices represent an arbitrary density matrix \cite{Fano}. Moreover, the Fano representation is convenient for numerical simulation of the dynamics also. By calculating the eigenvalues of the Jacobian in the Fano representation for each numerically found cycle, we concluded that among them the only stable cycle is the length two cycle $\{\rho_{1},\rho_{2}\}$, where
\begin{eqnarray}
\label{eq:SeparableMixed2QAttract}
\rho_{1}&=&\frac{1}{2}\left(|00\rangle\langle 00|+|11\rangle\langle 11|\right)
,\;\,\\
\rho_{2}&=&\frac{1}{4}\left(|00\rangle\langle 00|+|11\rangle\langle 11|+\left(|01\rangle+|10\rangle\right)\left(\langle 01|+\langle 10|\right)\right) .  \nonumber
\end{eqnarray}
The same calculation for the cycles known from the pure initial state case indicated their stability against perturbation by arbitrary mixed states.

\begin{figure}
\begin{center}
\includegraphics[width=0.42 \textwidth]{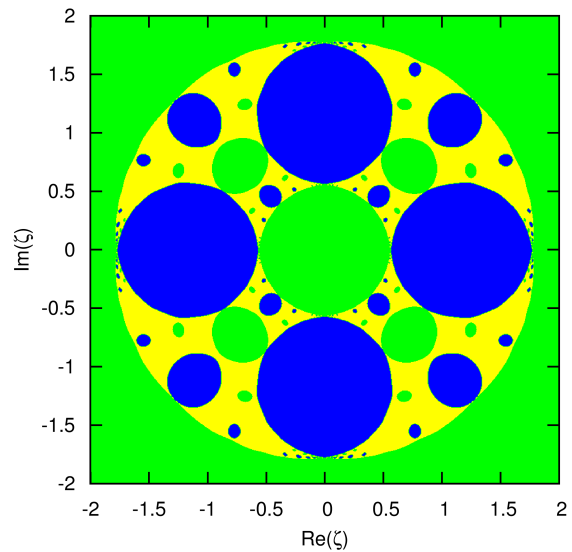}
\caption{ Convergence of mixed initial states $\rho(\lambda,\zeta)$ of
  Eq.~(\ref{eq:Werner}) to the three limiting cycles as a function of
  the complex parameter $\zeta$, at $\lambda=0.75$. The blue and green
  colours denote convergence to the same cycles as on Fig.~\ref{fig2},
  while the yellow colour denotes convergence to the separable mixed
  cycle in Eq.~(\ref{eq:SeparableMixed2QAttract}). As $\lambda$ is
  increased from 0 to 1, the corresponding convergence plots gradually
  turn from that on Fig.~\ref{fig2} to uniform yellow.}
\label{fig3}
\end{center}
\end{figure}

In Fig. \ref{fig3} we show the convergence towards the stable cycles of the mixed state dynamics.
The third, mixed stable cycle $\{\rho_{1},\rho_{2}\}$ denoted by yellow color washes out the fine structure of the pure state picture. Islands of purification towards both fully entangled and completely separable states remain visible even for the value $\lambda=0.75$ of the mixing parameter. Understanding the structure of convergent areas requires further studies. An interesting question is, whether the number of the islands purifying to entangled states is finite or it possesses a fractal structure.

{\em Conclusions---}We have demonstrated that entanglement in a quantum system can evolve truly chaotically, exhibiting sensitivity to the initial condition. 
Our results give an insight into the properties of the general pure state dynamics of a protocol that is described by a non-linear dynamical map in three complex variables. We have found that extending the space of initial states to a special class of mixed states, a new, mixed attracting cycle appears. It would require further studies to decide, whether adding other types of noise to the initial state would introduce additional attractors.

We acknowledge the financial support by MSM 6840770039,
M\v SMT LC 06002 and the Czech-Hungarian cooperation project
(KONTAKT, CZ-11/2009) and by the Hungarian
Scientific Research Fund (OTKA) under Contract No. K83858.


\begin{thebibliography}{20}
\bibitem{SchmidtBasisNote} The unique measure of entanglement for a
  bipartite pure state can be determined by writing it down in its
  Schmidt basis. For the states $|\Psi(\zeta)\rangle$ this basis
  coincides with the computational basis.
\end{thebibliography}
\end{document}